%% file: main.tex
\documentclass[sigconf]{acmart}

\usepackage{amsmath,amsfonts}
\usepackage{bm}
\usepackage{graphicx} 
\usepackage{caption}
\usepackage{subcaption}
\usepackage{natbib}  
\usepackage{caption} 
\usepackage{textcomp}
\usepackage[shortlabels]{enumitem}
\usepackage{balance}
\usepackage{comment}

\usepackage{url}
\usepackage{color}

\usepackage{algorithm}
\usepackage{algpseudocode}
\usepackage{listings}
\usepackage{multirow}
\usepackage{titlesec}
\usepackage{etoolbox}
\usepackage{blindtext}
\usepackage{wrapfig}
\usepackage{tcolorbox}

\AtBeginDocument{%
	\providecommand\BibTeX{{%
			\normalfont B\kern-0.5em{\scshape i\kern-0.25em b}\kern-0.8em\TeX}}}

\newcommand{\inlinedComment}[2]
{\textcolor{#1}{\small\textbf{#2}}}
\newcommand{\n}[1]{}
\newcommand{\lx}[1]{}
\newcommand{\yu}[1]{}

\copyrightyear{2021}
\acmYear{2021}
\setcopyright{acmcopyright}\acmConference[SIGIR '21]{Proceedings of the 44th International ACM SIGIR Conference on Research and Development in Information Retrieval}{July 11--15, 2021}{Virtual Event, Canada}
\acmBooktitle{Proceedings of the 44th International ACM SIGIR Conference on Research and Development in Information Retrieval (SIGIR '21), July 11--15, 2021, Virtual Event, Canada}
\acmPrice{15.00}
\acmDOI{10.1145/3404835.3462840}
\acmISBN{978-1-4503-8037-9/21/07}


\begin{document}
   \fancyhead{}
   \title{Self-Supervised Contrastive Learning for Code Retrieval and\\ Summarization via Semantic-Preserving Transformations}
	
	\author{Nghi D. Q. Bui}
	\thanks{This work was mostly done when the first author was working in the School of Computing and Information Systems, Singapore Management University.}
	\authornotemark[1]
	\affiliation{
	\institution{Trustworthy Software Engineering \& Open Source Lab}
\institution{Huawei Ireland Research Center}}
	\email{nghi.bui@huawei.com}
	
	\author{Yijun Yu}
	\affiliation{%
		\institution{Trustworthy Software Engineering \& Open Source Lab}
		\institution{Huawei Ireland Research Center \& The Open University, UK}}
	\email{yijun.yu@huawei.com}
		
	\author{Lingxiao Jiang}
	\affiliation{
		\institution{School of Computing \& \\Information Systems} \institution{Singapore Management University}}
	\email{lxjiang@smu.edu.sg}

%
%
	
    \begin{abstract}
   	\input{abstract}
    \end{abstract}
    \maketitle

    \input{body}

\balance
\bibliographystyle{ACM-Reference-Format}
\bibliography{references}

\end{document}

%% file: abstract.tex

%
We propose {\it Corder}, a self-supervised contrastive learning framework for source code model. Corder is designed to alleviate the need of labeled data for code retrieval and code summarization tasks. 
The pre-trained model of Corder can be used in two ways: (1) it can produce vector representation of code which can be applied to code retrieval tasks that do not have labeled data; (2) it can be used in a fine-tuning process for tasks that might still require label data such as code summarization. 
The key innovation is that we train the source code model by asking it to recognize similar and dissimilar code snippets through a \textit{contrastive learning objective}. To do so, we use a set of semantic-preserving transformation operators to generate code snippets that are syntactically diverse but semantically equivalent. %
Through extensive experiments, we have shown that the code models pretrained by Corder substantially outperform the other baselines for code-to-code retrieval, text-to-code retrieval, and code-to-text summarization tasks.

%% file: body.tex
\section{Introduction} \label{sec:introduction}
\input{introduction}

%

\vspace{-7pt}
\section{Related Work}
\label{sec:related}
\input{related}


\section{Approach}
\label{sec:approach}
\input{approach}

\section{Use Cases}
\label{sec:use_case}
\input{use_case}

\section{Empirical Evaluation}
\label{sec:evaluation}
\input{evaluation}

\vspace{-7pt}
\section{Analysis and Ablation Study}
\label{sec:analysis}
\input{analysis}

\vspace{-7pt}
\section{Conclusion} \label{sec:conclusion}
\input{conclusion}

\section*{Acknowledgements}
\label{sec:ack}
\input{ack}

%% file: introduction.tex
Deep learning models for code have been found useful in many software engineering tasks, such as predicting bugs~\cite{Yang2015,li2017software,zhou2019devign, he2021learning}, translating programs~\cite{chen2018tree, Gu2017}, classifying program functionality~\cite{Nix2017,dahl2013large}, searching code \cite{gu2018deep,kim2018facoy,Sachdev2018}, generating comments from code \cite{hu2018deep,Wan2018,Alon2019, wang2021automatic}, etc. These tasks can be seen as {\tt code retrieval} where code could be either the documents to be found or the query to search.
To build indices or models of source code for retrieval, a key step is to extract patterns of semantically equivalent or non-equivalent code snippets in large quantities. This is challenging for tremendous human efforts to collect and label code snippets.
To overcome this challenge, one can depend on heuristics to label the code snippets automatically, e.g. by using test cases to compare programs~\cite{Massalin87}. The downside of such an approach is the extra costs associated with code execution, which may not always be possible. Another way is to collect free source code on hosting platforms such as Github, and extract the snippets that share similar comments~\cite{hu2018deep}, method names~\cite{Alon2019}, or code documents, and then treat such snippets as semantically equivalent~\cite{husain2019codesearchnet}. The drawback to this heuristic is that it can add a lot of noise because not all code snippets of identical comments, method names, or documents are indeed semantically equivalent.
For example, \citet{kang2019assessing}
show that when pre-trained specifically for the method-name prediction task, the pre-trained Code2vec~\cite{Alon2019} model does not perform well for other code modeling tasks.
\citet{jiang2019machine} perform further analysis to show the reason that methods with similar names are not necessarily semantically equivalent, which explains the poor transfer learning results of \citet{kang2019assessing} on Code2vec since the model is forced to learn incorrect patterns of code.

\begin{figure}[t]
	\centering
	\includegraphics[width=1.0\columnwidth]{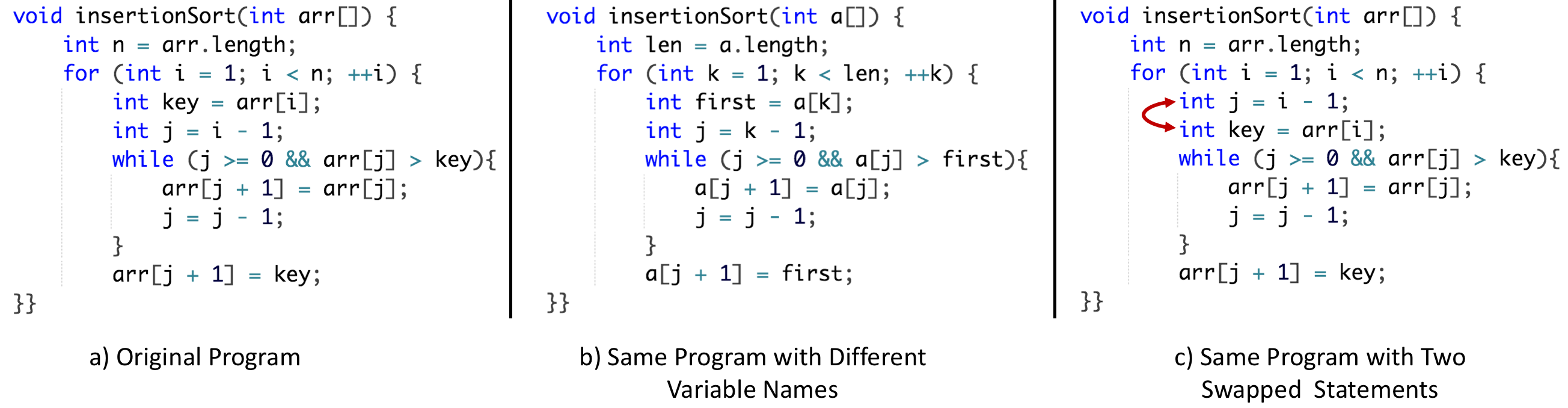}
	\caption{An Example of Semantically Equivalent Programs}
	\label{figure:intuition}
\end{figure}

To address these limitations, we develop Corder, a contrastive learning framework for code model that trains neural networks to identify semantically equivalent code snippets from a large set of transformed code. Essentially, the main goal of Corder is to invent a {\em pretext task} that enables the training of neural networks without the need for imprecise heuristics for identifying semantically equivalent programs. 
The pretext task that we implement here is called \textit{instance discrimination}, which is related to recent work of~\citet{chen2020simple}: a neural network is asked to discriminate instances of code snippets that are semantically equivalent from the instances of code snippets that are dissimilar.
In the end, the neural model is trained with the knowledge of how different instances of code snippets should be close to each other in a vector space depending on how likely they are semantically equivalent. The key idea is to leverage program transformation techniques that transform a code snippet into different versions of itself. Although syntactically different from the originals, these transformed programs are semantically equivalent. Figure~\ref{figure:intuition} shows an example of the transformed programs: Figure~\ref{figure:intuition}b shows a snippet semantically equivalent to the snippet in Figure~\ref{figure:intuition}a, with variables renamed. The snippet in Figure~\ref{figure:intuition}c is another transformed version of Figure~\ref{figure:intuition}a, with two independent statements swapped. The goal is then to train the neural network with these semantically equivalent snippets and ensure they are embedded closely in the vector space.

Corder uses the \textit{contrastive learning} methods that have been used in the self-supervised learning setting. The objective of contrastive learning is to simultaneously maximize the agreement between the differently transformed snippets of the same original snippet and minimize the agreement between the transformed snippets of different snippets. Updating the parameters of a neural network using this contrastive learning objective causes the representations of semantically equivalent snippets to be close to each other, while representations of non-similar snippets to be far apart. Once the model has been trained on our pretext task with the contrastive learning objective~\footnote{We call this the Corder pretext task.}, it can be used in two ways.
First, since it has trained a neural network encoder
(which is a part of the end-to-end learning process and can be instantiated with different encoders) 
and can be used to produce the representations of any source code. The vector representations of source code can be useful in many ways for code retrieval.
Second, the pre-trained model can be fine-tuned with a small amount of labelled data to achieve good performance for other downstream tasks, such as code summarization. In this work, we consider three concrete tasks that can leverage our pre-trained model, namely: code-to-code retrieval, text-to-code retrieval, and code summarization (which is code-to-text).
We have trained different
Corder instances with different encoders on large-scale Java datasets to evaluate their effectiveness for the tasks. 

To summarize, our major contributions are as follows:
\begin{itemize}[leftmargin=*]
	\item We explore a novel perspective of learning source code models from unlabeled data. Unlike existing work that uses imprecise heuristics to produce labelled data for training, we adapt program transformation techniques to generate precise semantically-equivalent code snippets for training. To the best of our knowledge, we are the first to use the program transformation technique for self-supervised learning of source code models.
	\item We develop Corder, a self-supervised contrastive learning framework, to identify semantically-equivalent code snippets generated from program transformation operators.
	\item We conduct extensive evaluations to demonstrate that the Corder pretext task is better than others in learning source code models in two ways: (1) we use the pre-trained models to produce vector representations of code and apply such representations in the {\it unsupervised} code-to-code retrieval task. The results show that any neural network encoder trained on the Corder pretext task outperforms the same encoders trained on other pretext tasks with a significant margin. Moreover, our technique outperforms other baselines that were designed specifically for code-to-code retrieval, such as FaCoy~\cite{kim2018facoy} significantly; (2) we use the pre-trained models in a fine-tuning process for {\it supervised} code modeling tasks, such as text-to-code retrieval and code summarization. The results show that our pre-trained models on the Corder pretext task perform better than training the code models from scratch and other pretext tasks, by a large margin.
\end{itemize}

%% file: related.tex

\paragraph{Self-Supervised Learning} has made tremendous strides in the field of visual learning ~\cite{mahendran2018cross, gidaris2018unsupervised, zhang2016colorful, korbar2018cooperative, kim2019self, fernando2017self}, and for quite some time in the field of natural language processing~\cite{mikolov2013distributed, le2014distributed, kiros2015skip, devlin2018bert, wu2019self, kong2019mutual, wang2019self, shi2020simple}. Such techniques allow for neural network training without the need for human labels. Typically, a self-supervised learning technique reformulates an unsupervised learning problem as one that is supervised by {\em generating virtual labels automatically from existing (unlabeled) data}. \textit{Contrastive learning} has emerged as a new paradigm that brings together multiple forms of supervised learning problem as the task to compare similar and dissimilar items, such as siamese neural networks~\cite{bromley1994signature}, contrastive predictive coding ~\cite{oord2018representation}, triplet loss~\cite{schroff2015facenet}. Contrastive learning methods specifically minimize the distance between similar data (positives) representations and maximize the distance between dissimilar data (negatives). 

\paragraph{Deep Learning Models of Code}: There has been a huge interest in applying deep learning techniques for software engineering tasks such as program functionality classification~\cite{mou2016convolutional,zhang2019novel,bui2017cross, bui2019bilateral}, bug localization~\cite{pradel2018deepbugs, gupta2019neural,cheng2021deepwukong, jarman2021legion}, code summarization~\cite{fernandes2018structured, ahmad2020transformer, svyatkovskiy2020intellicode}, code clone detection~\cite{zhang2019novel, bui2020infercode}, program refactoring~\cite{hu2018deep}, program translation~\cite{chen2018tree, bui2019sar}, and code synthesis~\cite{brockschmidt2018generative, alon2020structural}.
\citet{Allamanis2018} extend ASTs to graphs by adding a variety of code dependencies as edges among tree nodes, intended to represent code semantics, and apply Gated Graph Neural Networks (GGNN)~\cite{Li2016} to learn the graphs;
Code2vec~\cite{Alon2019}, Code2seq~\cite{alon2018code2seq}, and ASTNN~\cite{zhang2019novel} are designed based on splitting ASTs into smaller ones, either as a bag of path-contexts or as flattened subtrees representing individual statements. They use various kinds of Recurrent Neural Network (RNN) to learn such code representations. Surveys on code embeddings~\cite{Ingram2018,chen2019literature} present evidence to show that there is a \textbf{\textit{strong need to alleviate the requirement of labeled data for code modeling}} and encourage the community to invest more effort into the methods of learning source code with unlabeled data. Unfortunately, there is little effort towards designing the source code model with unlabeled data: \citet{yasunaga2020graph} presents a self-supervised learning paradigm for program repair, but it is designed specifically for program repair only. There are methods, such as~\cite{hussain2020deep,feng2020codebert} that perform pretraining source code data on natural language model (BERT, RNN, LSTM), but they simply train the code tokens similar to the way pretrained language models on text do, so they miss a lot of information about syntactical and semantic features of code that could have been extracted from program analysis. 

%% file: approach.tex

\begin{figure}[t]
	\centering
	\includegraphics[width=1.06\columnwidth]{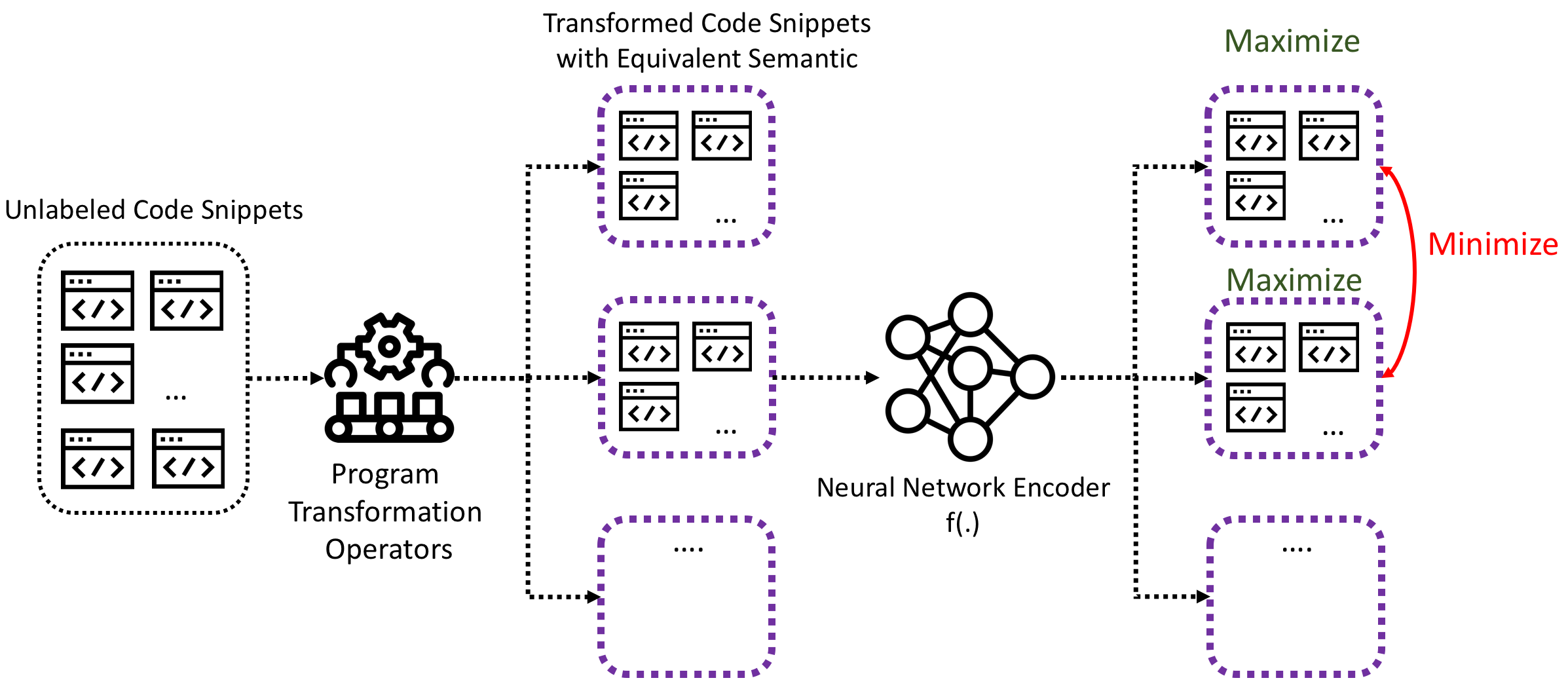}
	\caption{Overview of Corder pretext task. Unlabeled code snippets from a large codebase go through a program transformation module. Snippets in the purple dashed box are transformed snippets from the same original snippet. The goal is to maximize the similarity of the snippets in the same purple dashed box and minimize the similarity of snippets across different boxes}
	
	\label{figure:overview}
\end{figure}

\subsection{Approach Overview}

Figure~\ref{figure:overview} presents an overview of our approach. Overall, this framework comprises the following three major components.



\begin{itemize}[nosep,leftmargin=*]
	\item \textbf{A program transformation module} that transforms a given code snippet $p$, resulting in two transformed programs of the code snippets, denoted $\tilde{p}_i$ and $\tilde{p}_j$.
	\item \textbf{A neural network encoder} $f(\cdot)$ that receives an intermediate representation of a code snippet (such as Abstract Syntax Tree (AST)) and map it into a vector representation. In our case, it should map $\tilde{p}_i$ and $\tilde{p}_j$ into two code vectors $v_i$ and $v_j$, respectively.
	\item \textbf{A contrastive loss function} is defined for the contrastive learning task. 
	Given a set $\{\tilde{p}_k\}$ containing a positive (semantically similar) pair of examples $p_i$ and $\tilde{p}_j$, the \textit{contrastive prediction task} aims to identify $\tilde{p}_j$ in $\{\tilde{p}_k\}_{k\ne i}$ for a given $p_i$.
\end{itemize}

\begin{algorithm}[!t]
	\label{algo_1}
	\small
	\caption{Corder's learning algorithm}\label{algo:vts_routing}
	\begin{algorithmic}[1]
		\State \textbf{input:} batch size $N$, encoder $f$, set of transformation operators $\mathcal{T}$.
		\For{sampled minibatch $\{p_k\}_{k=1}^N$}
		\State \textbf{for all} $k\in \{1, \ldots, N\}$ \textbf{do}
		\State $~~~~$draw two transformation operators $t \!\sim\!
		\mathcal{T}$, $t' \!\sim\! \mathcal{T}$
		\State $~~~~$\textcolor{gray}{\# the first transformation}
		\State $~~~~$$\tilde{p}_{i} = t(p_k)$
		\State $~~~~$$\bm v_{i} = f(\tilde{p}_{i})$  
		\State $~~~~$\textcolor{gray}{\# the second transformation} 
		\State $~~~~$$\tilde{p}_{j} = t'(p_k)$
		\State $~~~~$$\bm v_{j} = f(\tilde{p}_{j})$
		\State \textbf{end for}
		\State \textbf{for all} $i\in\{1, \ldots, 2N\}$ and $j\in\{1, \dots, 2N\}$ \textbf{do}
		\State $~~~~$ $s_{i,j} = \bm z_i^\top \bm z_j / (\lVert\bm z_i\rVert \lVert\bm z_j\rVert)$ \textcolor{gray}{~~~~~~~~\# pairwise similarity}\\
		\State \textbf{end for}
		\State \textbf{define} $\ell(i, j)$ \textbf{as}~ $\ell(i, j) \!=\! -\log \frac{\exp(s_{i,j})}{\sum_{k=1}^{2N} 1_{k \neq i}\exp(s_{i, k})}$ \\ 
		\State $\mathcal{L} = \frac{1}{2N} \sum_{k=1}^N \left[ \ell(2k\!-\!1, 2k) + \ell(2k, 2k\!-\!1)\right]$
		\State update networks $f$ to minimize $\mathcal{L}$
		\EndFor
		\State \textbf{return} encoder network $f(\cdot)$
	\end{algorithmic}
\end{algorithm}

\subsection{Approach Details}

With the above components, here we describe the Corder training process in the following steps. Also, a summarization of the proposed algorithm is depicted in Algorithm 1. 
\begin{itemize}[leftmargin=*]
	\item A mini-batch of $N$ samples is randomly selected from a large set of code snippets. Each code snippet $p$ in $N$ is applied with two different randomly selected transformation operators, resulting in 2N transformed code snippets:
	$\tilde{p}_i = t(p); \; \tilde{p}_j = t'(p); \; t,t'\sim\mathcal{T}$, where $p$ is the original code snippet, $\tilde{p}_i$ and $\tilde{p}_j$ are transformed code snippets by applying two transformation operators $t$ and $t'$ into $p$, respectively. $t$ and $t'$ are randomly chosen from a set of available operators $\mathcal{T}$. 
	\item Each of the transformed snippet $\tilde{p}_i$ and $\tilde{p}_j$ will be fed into the same encoder $f(\cdot)$ to get the embedding representations:
	$v_i = f(\tilde{p}_i) \;; v_j = f(\tilde{p}_j).$

	\item We use the Noise Contrastive Estimate (NCE) loss function~\cite{chen2020simple} to compute the loss.
	Let $\mathrm{sim}(\bm u,\bm v) = \frac{\bm u^\top \bm v} {\lVert\bm u\rVert \lVert\bm v\rVert}$ denote the dot product between $\ell_2$ normalized $\bm u$ and $\bm v$ (i.e. cosine similarity). Then the loss function for a pair of representations $(v_{i}, v_{j})$ is defined as
	$
	\ell(i,j) = -log\frac{exp(sim(v_{i}, v_{j}))}{\sum_{k=1}^{2N} \mathbf{1}_{k \neq i}\exp(\mathrm{sim}(v_i, v_k))}
	$, where $\mathbf{1}_{k \neq a} \in \{ 0,  1\}$ is an indicator function evaluating to $1$ iff $k \neq i$. Noted that for a given positive pair, the other $2(N-1)$ transformed code snippets are treated as negative samples. We calculate the loss for the same pair a second time as well where the positions of the samples are interchanged.
	The final loss is computed across all pairs in a mini-batch can be written as:
	$
	L = \frac{1}{2N}\sum_{k=1}^N[\ell(2k-1, 2k) + \ell(2k,2k-1)].
	$

\end{itemize}

\subsubsection{Program Transformation Operators.}
\label{sec:transformation_operators}

\begin{figure}[t]
	\centering
	\includegraphics[width=1.0\columnwidth]{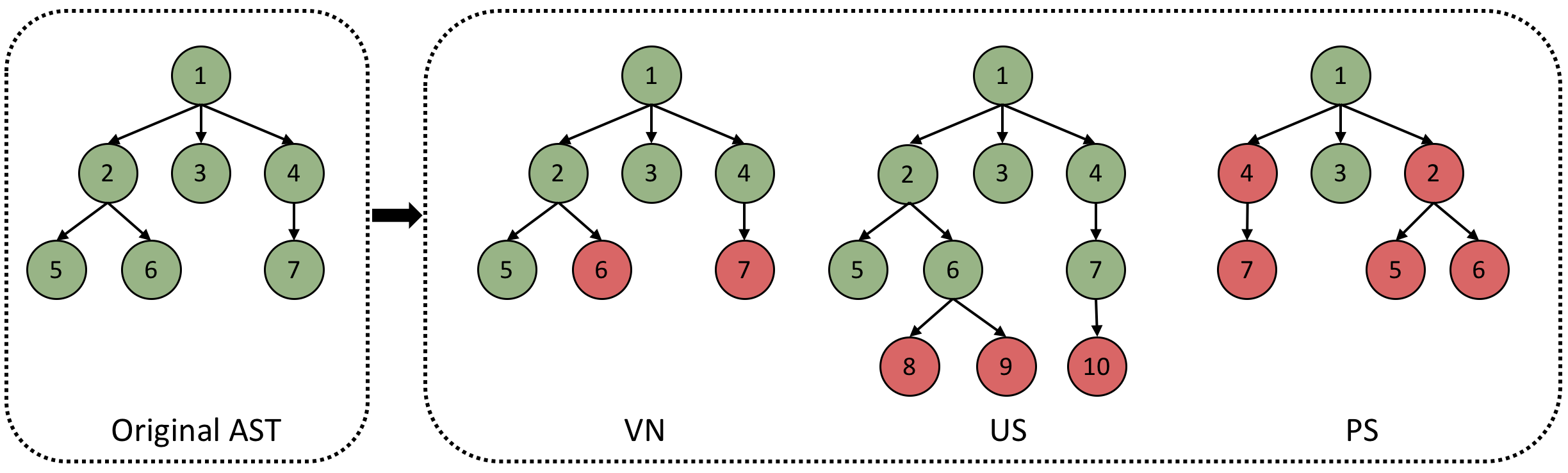}
	\caption{Example of how the AST structure is changed with different transformation operators}
	\label{figure:transformation}
\end{figure}

Our key idea to enable the encoder to learn a set of diverse code features without the need for labeled data is to generate multiple versions of a program without changing its semantics.
To do so, we 
apply a set of semantic-preserving program transformation operators to generate such different variants.
There are many methods for transforming the code~\cite{rabin2020generalizability}, and the more sophisticated a transformation is, in principle, the better the encoder can learn the essential semantic of the code.
In this work, we mainly apply the following transformations:
variable renaming, adding dead code (unused statements), permuting statements, loop exchange, and switch to {\tt if}, to reflect different ways to change the structure of the Abstract Syntax Tree (AST). 
We evaluate how different transformations can effect the performance of the encoder in Section~\ref{impact}.

\begin{itemize}[leftmargin=*]
	\item \textbf{Variable Renaming (VN)} is a refactoring method that renames a variable, where the new name of the variable is taken randomly from a set of variable vocabulary in the training set. Noted that each time this operator is applied to the same program, the variable names are renamed differently. This operator does not change the structure of the AST representation of the code, it only changes the textual information, which is a feature of a node in the AST. 
	\item \textbf{Unused Statement (US)} 
	is to insert dead code fragments, such as unused statement(s) to a randomly selected basic block in the code. We traverse the AST to identify the blocks and randomly select one block to insert predefined dead code fragments into it. This operator will add more nodes to the AST. To diversify the set of transformed programs, we prepare a large set of unused statement(s). When the operator is applied, random statements in the set is selected to added into the code block, i.e., a transformed snippet is different each time we apply the same operator. 
	\item \textbf{Permutation of Statements (PS)} is to swap two statements that have no dependency on each other in a basic block in the code. We traverse the AST and analyze the data dependency to extract all of the possible pairs of swap-able statements. If a program only contains one such pair, it will generate the same output every time we apply the operator, otherwise, the output will be different. 
	\item \textbf{Loop Exchange (LX)} replaces for loops with while loops or vice versa. We traverse the AST to identify the node the defines the for loop (or the while loop) then replace one with another with modifications on the initialization, the condition, and the afterthought.
	\lx{is this used in the evaluation?}
	\item \textbf{Switch to If (SF)} replaces a switch statement in the method with its equivalent if statement. We traverse the AST to identify a switch statement, then extract the subtree of each case statement of the switch and assign it to a new if statement.
	
\end{itemize}

Each of the transformation operators above is designed to change the structure representation of the source code differently. For example, with VR, we want the NN to understand that even the change in textual information does not affect the semantic meaning of the source code, inspired by a recent finding of ~\citet{zhang2020generating}. It is suggested that the source code model should be equipped with adversarial examples of token changes to make the model become more robust. With US, we want the NN still to learn how to catch the similarity between two similar programs even though the number of nodes in the tree structure has increased. With PS, the operator does not add nodes into the AST but it will alter the location of the subtrees in the AST, we want the NN to be able to detect the two similar trees even if the positions of the subtrees have changed. Figure~\ref{figure:transformation} illustrates how the AST structure changes with the corresponding transformation operator.

\subsubsection{Neural Network Encoder for Source Code}

The neural network can also be called as an {\em encoder}, written as a function $f(\cdot)$. The encoder receives the intermediate representation (IR) of code and maps it into a code vector embedding $\vec{v}$ (usually a combination of various kinds of code elements), then $\vec{v}$ can be fed into the next layer(s) of a learning system and trained for an objective function of the specific task of the learning system. The choice of the encoder depends mostly on the task and we will rely on previous work to choose suitable encoders for a particular task, which will be presented in Section~\ref{sec:evaluation}.

%% file: use_case.tex

%

We present three tasks (code-to-code retrieval, text-to-code retrieval, and code summarization) to make good uses of the pre-trained Corder models in two ways. 


.


\vspace{-10pt}
\subsection{Using the Pre-trained Encoders to Produce Code Vectors for  Downstream Task}

The first way to use pre-trained encoders from our Corder pretext task is to use such encoders to produce the vector representations of code. Then the representations can be applicable for a downstream task, such as code-to-code retrieval.

\subsubsection{Code-to-Code Retrieval}
\begin{figure}[!t]
	\centering
	\includegraphics[width=1.0\columnwidth]{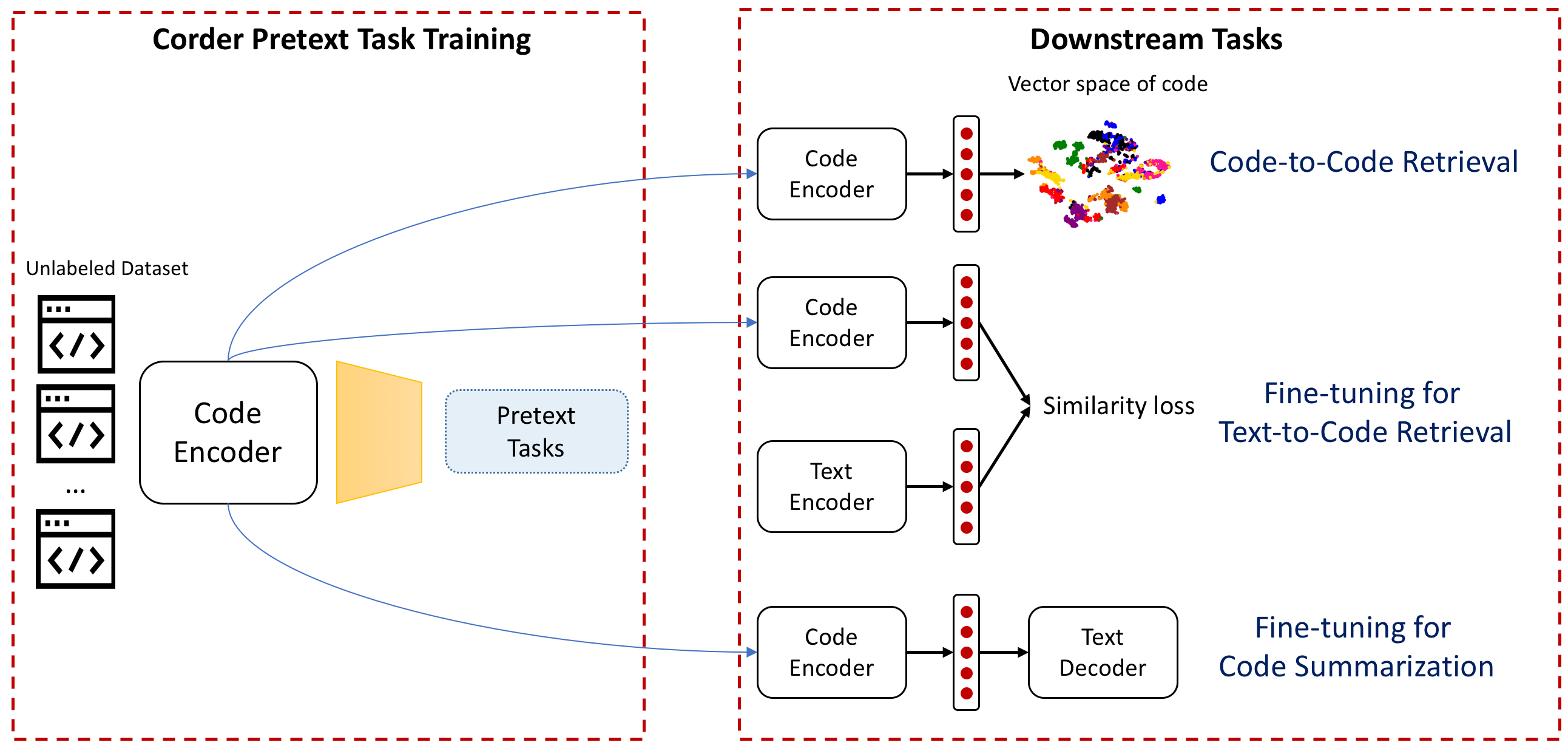}
	\caption{Process on how Corder pre-trained model can be applied in different downstream tasks}
	\label{figure:tasks}
\end{figure}

\textit{Code-to-code} search is useful for developers to find other code in a large codebase that is similar to a given code query. Most of the work that is designed for code-to-code retrieval, such as Facoy~\cite{kim2018facoy}, Krugle~\cite{Krugle} is based on the simple text mining approach or traditional code clone detection method. These techniques required tremendous effort of handcraft feature engineering to extract good features of code. In our case, we adapt pre-trained source code encoders from the Corder pretext task to map any code snippet into a vector representation, then we perform the retrieval task based on the vectors (see Figure~\ref{figure:tasks}, Code-to-Code Retrieval). Assume that we have a large codebase of snippets, we used the pre-trained encoders to map the whole codebase into representations. Then for a given code snippet as a query, we map such query into vector representation too. 
Then, one can find the top-k nearest neighbors of such query in the vector space, using cosine similarity
as the distance metric, and finally can retrieve the list of candidate snippets. These snippets are supposed to be semantical equivalent to the query. 


\subsection{Fine-Tuning the Encoders for Supervised Learning Downstream Tasks}
A  paradigm to make good use of a large amount of unlabeled data is self-supervised pre-training followed by a supervised fine-tuning~\cite{hinton2006fast,chen2020simple}, which reuses parts (or all) of a  trained neural network on a  certain task and continue to train it or simply using the embedding output for other tasks. Such  fine-tuning  processes usually have the benefits of (1) speeding up the training as one  does not need to train the model from randomly initialized weights and (2) improving the generalizability of the downstream model even when there are only small datasets with labels.
As shown in  Figure~\ref{figure:tasks}, the encoder is used as a pre-trained model in which the weights resulting from the Corder pretext task are  transferred to initialize the model of the downstream supervised learning tasks.

\subsubsection{Text-to-Code Retrieval}
This task is to, given a natural language as the query, the objective is to find the most semantically related code snippets from a collection of codes~\cite{gu2018deep, husain2019codesearchnet}. Note that this is different from the \textit{code-to-code} retrieval problem, in which the query is a code snippet. The deep learning framework used in the literature for this task is to construct a bilateral neural network structure, which consists of two encoders, one is a \textit{natural language encoder} (such as BERT, RNN, LSTM) to encode text into text embedding, the other is a \textit{source code encoder} to encode an immediate source code representation into the code embedding ~\cite{husain2019codesearchnet, gu2018deep}. Then, from text embedding and code embedding, a mapping function is used to push the text and the code to be similar to the vector space, called a shared embedding between the code and the text. In the retrieval process, the text description is given, and we use its embedding to retrieve all the embeddings of the code snippets that are closest to the text embedding. In the fine-tuning process, the source code encoder that has been pre-trained on the Corder pretext task will be used to initialize for the source code encoder, the parameters of the text encoder will be initialized randomly.

\subsubsection{Code Summarization}

The purpose of this task is to predict a concise text description of the functionality of the method given its source code ~\cite{BaroneS17}. Such descriptions typically appear as documentation of methods (e.g. "docstrings" in Python or "JavaDocs" in Java). This task can be modeled as a translation task where the aim is to translate a source code snippet into a sequence of text. As such, the encoder-decoder model, such as seq2seq~\cite{sutskever2014sequence} is usually used in the literature for this task. In our case, the encoder can be any code modeling technique, such as TBCNN~\cite{mou2016convolutional}, Code2vec~\cite{Alon2019}, LSTM or Transformer. In the fine-tuning process, the source code encoder that has been pre-trained on the Corder pretext task will be used to initialize for the source code encoder.

%% file: evaluation.tex
\subsection{Settings}
\label{eval:settings}
\subsubsection{Data Preparation}
As presented, we will perform the evaluation on three tasks, namely, code-to-code search, text-to-code search, and code summarization. 
We used the JavaSmall and JavaMed datasets that have been widely used recently for code modeling tasks~\cite{Alon2019, alon2018code2seq}. JavaSmall is a dataset of 11 relatively large Java projects from GitHub, which contains about 700k examples.
JavaMed is a dataset of 1000 top-starred Java projects from GitHub which contains about 4M examples.



Then, we parse all the snippets into ASTs using SrcML~\cite{collard2013srcml}. We also perform the transformation on all of the ASTs to get the transformed ASTs based on the transformation operators described in Section~\ref{sec:transformation_operators}, having the ASTs as well as the transformed ASTs. It should be noted that SrcML is a universal AST system, which means that it uses the same AST representations for multiple languages (Java, C\#, C++, C). This enables the model training on each of the languages once and they can be used in other languages.  Another note is that the two datasets are not the ones used for evaluation purposes; they are only for the purpose of training the Corder pretext task on different encoders. We describe the evaluation datasets used for each of the tasks separately in each of the subsections.

\subsubsection{Encoders}

We choose a few well-known AST-based code modeling techniques as the encoder $f(\cdot)$, which are Code2vec~\cite{Alon2019}, TBCNN~\cite{mou2016convolutional},
We also include two token-based techniques by treating source code simply as sequences of tokens and using a neural machine translation (NMT) baseline, i.e. a 2-layer Bi-LSTM, and the Transformer~\cite{vaswani2017attention}. A common setting used among all these techniques is that they all utilize both node type and token information to initialize a node in ASTs.

We set both the dimensionality of type embeddings and text embeddings to 128. Note that we try our best to make the baselines as strong as possible by choosing the hyper-parameters above as the ``optimal settings'' according to their papers or code.
Specifically, for Code2vec~\cite{Alon2019}\footnote{\scriptsize \url{https://github.com/tech-srl/code2vec}} and Code2seq~\cite{alon2018code2seq}\footnote{\scriptsize \url{https://github.com/tech-srl/code2seq}}, since Code2seq is a follow-up work of Code2vec (different in the decoder layer to predict the sequence), we follow the settings in Code2seq to set the size of each LSTM encoders for ASTs to 128 and the size of LSTM decoder to 320. We set the number of paths sampled for each AST to 200 as suggested, since increasing this parameter does not improve the performance. TBCNN~\cite{mou2016convolutional} uses a tree-based convolutional layer with three weight matrices serving as model parameters to accumulate children's information to the parent, each will have the shape of 128 x 128. We set the number of convolutional steps to 8. For Transformer~\cite{vaswani2017attention}, we choose to set the number of layers to 5 and the attention dimension size to 128. Finally, for the 2-layer Bi-LSTM, we followed the strategy from \citet{alon2018code2seq} by assigning the token embedding size to 128, the size of the hidden unit in the encoder to 128, and the default hyperparameters of OpenNMT~\cite{klein2017opennmt}.

\subsubsection{Research Questions}
We want to answer two research questions specifically through the evaluations:
\begin{enumerate}[(1),nosep,leftmargin=*]
	\item Are the code vectors generated by the pre-trained modesl (with various encoders) useful in the unsupervised space searching task (code-to-code search in particular)?
	\item Can the pre-trained models be used in a fine-tuning process to improve the performance of the downstream supervised models for text-to-code search and code summarization without training the models from scratch?
\end{enumerate}

\subsection{Using Pre-trained Encoders to Produce Code Representations for Code-to-Code Retrieval} 
\subsubsection{Datasets, Metrics, and Baselines}
Given a code snippet as the input, the task aims to find the most semantically related code from a collection of candidate codes. 
The datasets we used to evaluate for this task are:

\begin{itemize}[leftmargin=*]
	\item OJ dataset~\cite{mou2016convolutional} contains 52000 C programs with 104 classes, which results in 500 programs per class. Since the dataset is for C++, we translate the whole dataset with the C++ to Java Converter~\footnote{\url{https://www.tangiblesoftwaresolutions.com/product_details/cplusplus_to_java_converter_details.html}} to make the language of the evaluation dataset aligned with the pretrained models for Java (see Section~\ref{eval:settings}). Then we use the data that has been translated to Java for evaluation.
	\item BigCloneBench (BCB) dataset~\cite{svajlenko2015evaluating} contains 25,000 Java projects, cover 10 functionalities and including 6,000,000 true clone pairs and 260,000 false clone pairs. This dataset has been widely used for code clone detection task.
\end{itemize}

The OJ and BigCloneBench datasets have been widely used for the code clone detection task. The code clone detection task is to detect semantically duplicated (or similar) code snippets in a large codebase. Thus these datasets are also suitable for the code-to-code retrieval task, with the aim to find the most semantically related codes given the code snippet as the query. 

We randomly select 50 programs per class as the query, so that the total number of queries is 5200 for 104 classes. For each of the queries, we want to retrieve all of the semantically similar code snippets, which are the programs in the same class of the query.  With OJ, each query can have multiple relevant results, so that we use Mean Average Precision (MAP) as the metric to evaluate for the code-to-code search on the OJ dataset. Mean average precision for a set of queries is the mean of the average precision scores for each query, which can be calculated as
$
MAP = \frac{\sum_{q=1}^{Q}AveP(q)}{Q}
$, where $Q$ is the number of queries in the set and $AveP(q)$ is the average precision for a given query q.

For the BCB dataset, since the size of the dataset is large, we reduce the size by randomly select 50,000 sample clone pairs and 50,000 samples none clone pairs and evaluate within these pairs. Then within the clone pairs, we again randomly select 5000 pairs and pick one code snippet of a pair as the query. Let's denote a clone pair as $p=(c_{1}, c_{2})$, we pick $c_{1}$ as the query. For each of the query $c_{1}$, we want to retrieve the $c_{2}$, which is the snippet that is semantically identical to the query. With BCB, the assumption is that each query has only one relevant result so that we use Mean Reciprocal Rank (MRR) as the metric to evaluate for the task. Mean Reciprocal Rank is the average of the reciprocal ranks of results of a set of queries Q. The reciprocal rank of a query is the inverse of the rank of the first hit result. The higher the MRR value, the better the code search performance. MRR can be calculated as follows:
$
MRR = \frac{1}{|Q|}\sum_{q=1}^{|Q|}\frac{1}{rank_{i}}
$. Note that for both datasets, we limited the number of return results to 10. We use these baselines for the code-to-code retrieval task:

\begin{itemize}[leftmargin=*]
	\item Word2vec: the representation of the code snippet can be computed by simply calculate the average of the representations of all of the token in the snippet
	\item Doc2vec: we use Gensim~\footnote{\url{https://github.com/RaRe-Technologies/gensim}} to train the Doc2vec model on the JavaMed dataset and use the method provided by Gensim to infer the representation for a code snippet
	\item ElasticSearch: we treat the code snippet as a sequence and use the text tokenizer provided by ElasticSearch to index the code token and use ElasticSearch as a fuzzy text search baseline. 
	\item Facoy~\cite{kim2018facoy} is a search engine that is designed specifically for code-to-code search.
\end{itemize}

Besides the baselines above, we also want to see if our Corder pretext task performs better than the other pretext task for the same encoder. Among the encoders, the Transformer~\cite{vaswani2017attention} can be pre-trained with other pretext tasks, such as the masked language modeling, where a model uses the context words surrounding a [MASK] token to try to predict what the [MASK] word should be.
Code2vec~\cite{Alon2019} is also applicable for another pretext task, which is the method name prediction (MNP). The path encoder in Code2vec can encode the method body of a code snippet, then use the representation of the method body to predict the method name. With this, the Code2vec model can be pre-trained with MNP as a pretext task. The path encoder of the Code2vec for the method name prediction task can be reused to produce representation for any code snippet. For such reasons, we include 2 additional baselines, which are a pre-trained Transformer on the masked language model on the JavaMed dataset, and a pre-trained Code2vec on MNP on the JavaMed dataset.

\begin{table}[t]
	\centering
	\caption{Results of code-to-code retrieval. For BigCloneBench (BCB), the metric is MAP. For OJ, the metric is MRR }
	\label{tab:code_search}
	\fontsize{7.5}{8.6}\selectfont 
	\begin{tabular}{|c|c|c|c|c|}
		\hline
		\multirow{2}{*}{\textbf{Model}} & \multirow{2}{*}{\textbf{Pre-training}} & \multirow{2}{*}{\textbf{Dataset}} & \multicolumn{2}{c|}{\textbf{Performance}} \\ \cline{4-5} 
		&                                &                                   & \textbf{BCB (MRR)}   & \textbf{OJ (MAP)}  \\ \hline
		ElasticSearch                   & -                              & -                                 & 0.131                & 0.235              \\ \hline
		Word2Vec                        & -                              & JavaMed                           & 0.255                & 0.212              \\ \hline
		Doc2Vec                         & -                              & JavaMed                           & 0.289                & 0.334              \\ \hline
		FaCoy                           & -                              & JavaMed                           & 0.514                & 0.585              \\ \hline
		\multirow{4}{*}{Code2Vec}       & MNP                            & JavaSmall                         & 0.374                & 0.529              \\ \cline{2-5} 
		& MNP                            & JavaMed                           & 0.453                & 0.621              \\ \cline{2-5} 
		& Corder                         & JavaSmall                         & 0.561                & 0.631              \\ \cline{2-5} 
		& Corder                      & JavaMed                           & \textbf{0.633}                & \textbf{0.735}          \\ \hline
		\multirow{2}{*}{TBCNN}          & Corder                         & JavaSmall                         & 0.654                & 0.710              \\ \cline{2-5} 
		& Corder                         & JavaMed                           & \textbf{0.832 }               & \textbf{0.856 }             \\ \hline
		\multirow{2}{*}{Bi-LSTM}        & Corder                         & JavaSmall                         & 0.620                & 0.534              \\ \cline{2-5} 
		& Corder                         & JavaMed                           & \textbf{0.742}                & \textbf{0.690}              \\ \hline
		\multirow{4}{*}{Transformer}    & Masked LM                      & JavaSmall                         & 0.580                & 0.534              \\ \cline{2-5} 
		& Masked LM                      & JavaMed                           & 0.719                & 0.693              \\ \cline{2-5} 
		& Corder                         & JavaSmall                         & 0.640                & 0.720              \\ \cline{2-5} 
		& Corder                         & JavaMed                           & \textbf{0.825}                & \textbf{0.841}              \\ \hline
	\end{tabular}
\end{table}

\subsubsection{Results}

Table~\ref{tab:code_search} shows the results of the code-to-code retrieval task. The column "Pre-training" with different options, such as "Corder", "Masked LM", "MNP" means that an encoder is trained with a different pretext task. ElasticSearch, Word2vec, Doc2vec, and FaCoy are not applicable for such pretext tasks, hence "-" is used for these techniques in the column "Pre-training". The column "Dataset" means that an encoder is trained on a specific dataset, such as JavaSmall and JavaMed.

As one can see, ElasticSearch, an information retrieval approach, performs worst among the baselines. Word2vec and Doc2vec perform better but the results are still not so good. Code2vec and Bi-LSTM, when pre-training with the Corder process on the JavaMed, can perform better than FaCoy, a method designed specifically for code-to-code retrieval. Code2vec, 
when pre-training with the method name prediction (MNP) pretext task, performs much worse than the pre-training with the Corder pretext task. Transformer, when pre-training with the masked language model (Masked-LM) pretext task, performs much worse than the pre-training with the Corder pretext task. This shows that our proposed pretext task performs better than the other pretext tasks to train the representation of the source code.

\subsection{Fine-tuning Pre-trained Encoders for Text-to-code Retrieval} 

\subsubsection{Datasets, Metrics, and Baselines}
Given a natural language as input, the task aims to find the most semantically related code from a
collection of candidate codes. We use the dataset released by  DeepCS~\cite{gu2018deep}, which consists of approximately 16 million preprocessed Java methods and their corresponding docstrings.

For the metrics, we use Precision at k (Precision@k) and Mean Reciprocal Rank to evaluate this task.
Precision@k measures the percentage of relevant results in the top k returned results for each query. In our evaluations, it is calculated as follows:
$
Precision@k = \frac{\# relevant~results~in~the~top~k~results}{k}
$
. Precision@k is important because developers often inspect multiple
results of different usages to learn from. A better code retrieval
engine should allow developers to inspect less noisy results. The
higher the metric values, the better the code search performance.
We evaluate and Precision@k when the value of k is 1, 5,
and 10. These values reflect the typical sizes of results that users would inspect.

We choose to use the three methods presented in CodeSearchNet~\cite{husain2019codesearchnet} for the text-to-code retrieval models, which are:
neural bag-of-words, 2-layer BiLSTM, and Transformer. 
We also include Tree-based CNN (TBCNN)~\cite{mou2016convolutional} and Code2vec~\cite{Alon2019} which are  AST-based encoders that receive the AST representation of the code snippets as the input. 
We perform evaluations under two settings: (1) train from scratch and (2) fine-tune with a pre-trained model. In the second setting, each of the encoders will be pre-trained through the Corder pretext task, then the pre-trained encoder will be used for the fine-tuning process. We also include the pre-trained Code2vec model from the method name prediction (MNP) task to demonstrate that our Corder pretext task is better in a fine-tuning process than the pre-trained model from the MNP task.

\begin{table}[t]
	\centering
	\caption{Results of text-to-code retrieval}
	\label{tab:text_to_code_search}
	\fontsize{7.0}{8.5}\selectfont 
	\begin{tabular}{|c|c|c|c|c|c|c|} 
		\hline
		\textbf{Model}               & \textbf{Pre-training}  & \textbf{Dataset}   & \textbf{P@1}               & \textbf{P@5}               & \textbf{P@10}   & \textbf{MRR}     \\ 
		\hline
		NBow                         & -                      & -         & 0.394                      & 0.581                      & 0.603           & 0.384            \\ 
		\hline
		\multirow{3}{*}{Code2vec}    & -                      & -         & 0.406                      & 0.529                      & 0.564           & 0.395            \\ 
		\cline{2-7}
		& MNP                    & JavaSmall & 0.415                      & 0.538                      & 0.572           & 0.409            \\ 
		\cline{2-7}
		& MNP                    & JavaMed   & 0.435                      & 0.546                      & 0.583           & 0.420            \\ 
		\cline{2-7}
		& Corder                 & JavaSmall & 0.512                      & 0.578                      & 0.610           & 0.446            \\ 
		\cline{2-7}
		& Corder                 & JavaMed   & \textbf{0.549}             & \textbf{0.608}             & \textbf{0.625}  & \textbf{0.592}    \\ 
		\hline
		\multirow{3}{*}{TBCNN}       & -                      & -         & 0.506                      & 0.581                      & 0.632           & 0.551            \\ 
		\cline{2-7}
		& Corder                 & JavaSmall & 0.541                      & 0.620                      & 0.658           & 0.658            \\ 
		\cline{2-7}
		& Corder                 & JavaMed   & \textbf{0.640}             & \textbf{0.710}             & \textbf{0.758}  & \textbf{0.702}   \\ 
		\hline
		\multirow{3}{*}{Bi-LSTM}     & -                      & -         & 0.469                      & 0.540                      & 0.702           & 0.630            \\ 
		\cline{2-7}
		& Corder                 & JavaSmall & 0.532                      & 0.581                      & 0.723           & 0.619            \\ 
		\cline{2-7}
		& Corder                 & JavaMed   & \textbf{0.567}             & \textbf{0.639}             & \textbf{0.768}  & \textbf{0.661}   \\ 
		\hline
		\multirow{5}{*}{Transformer} & -                      & -         & 0.534                      & 0.653                      & 0.793           & 0.651            \\ 
		\cline{2-7}
		& Masked LM              & JavaSmall & \multicolumn{1}{l|}{0.567} & \multicolumn{1}{l|}{0.627} & 0.683           & 0.630            \\ 
		\cline{2-7}
		& Masked LM              & JavaMed   & 0.632                      & 0.710                      & 0.753           & 0.672            \\ 
		\cline{2-7}
		& Corder                 & JavaSmall & 0.604                      & 0.698                      & 0.845           & 0.687            \\ 
		\cline{2-7}
		& Corder                 & JavaMed   & \textbf{0.662}             & \textbf{0.756}             & \textbf{0.881}  & \textbf{0.728}   \\
		\hline
	\end{tabular}
\end{table}

\subsubsection{Results}

Table~\ref{tab:text_to_code_search} shows the performance of text-to-code retrieval task. The column "Pre-training" with different options, such as "-", "MNP", "Corder", means that an encoder is pre-trained on different pretext tasks. "-" means that there is no pretext task applied for the model and the model is trained from scratch. The column "Dataset", with different options, such as "-", "JavaSmall", "JavaMed", means that the encoder is pre-trained on a pretext task with a specific dataset.
There are 3 observations : (1) Corder pre-training task on any of the model improves the performance significantly; (2) pre-training on a larger dataset improves the results with a higher margin than pre-training on a smaller dataset, and (3) Corder pretext task for Code2vec performs better than the MNP task to fine-tune the model for text-to-code retrieval. 
\begin{table}[t!]
	\centering
	\caption{Results of code summarization}
	\label{tab:code_summarization}
	\fontsize{7.5}{8.0}\selectfont 
	
	\begin{tabular}{|c|c|c|c|}
		\hline
		\textbf{Model}               & \textbf{Pre-training} & \textbf{Dataset} & \textbf{BLEU}  \\ \hline
		MOSES                        & -                     & -                & 11.56          \\ \hline
		IR                           & -                     & -                & 14.32          \\ \hline
		\multirow{5}{*}{Code2seq}    & -                     & -                & 22.89          \\ \cline{2-4} 
		& MNP                   & JavaSmall        & 23.14          \\ \cline{2-4} 
		& MNP                   & JavaMed          & 24.20          \\ \cline{2-4} 
		& Corder                & JavaSmall        & 24.23          \\ \cline{2-4} 
		& Corder                & JavaMed          & \textbf{26.56} \\ \hline
		\multirow{3}{*}{TBCNN}       & -                     & -                & 21.15          \\ \cline{2-4} 
		& Corder                & JavaSmall        & 23.56          \\ \cline{2-4} 
		& Corder                & JavaMed          & \textbf{25.39} \\ \hline
		\multirow{3}{*}{Bi-LSTM}     & -                     & -                & 23.98          \\ \cline{2-4} 
		& Corder                & JavaSmall        & 24.21          \\ \cline{2-4} 
		& Corder                & JavaMed          & \textbf{25.50} \\ \hline
		\multirow{5}{*}{Transformer} & -                     & -                & 22.85          \\ \cline{2-4} 
		& Masked LM             & JavaSmall        & 23.10          \\ \cline{2-4} 
		& Masked LM             & JavaMed          & 24.78          \\ \cline{2-4} 
		& Corder                & JavaSmall        & 25.69          \\ \cline{2-4} 
		& Corder                & JavaMed          & \textbf{26.41} \\ \hline
	\end{tabular}
\end{table}
\subsection{Fine-tuning Pre-trained Encoders for Code Summarization} 
\subsubsection{Dataset, Metric, and Baselines}
For this task, we consider predicting a full natural language sentence given a short code snippet.
We also use the Java dataset provided by DeepCS~\cite{gu2018deep}, which consists of approximately 16 million preprocessed Java methods and their corresponding docstrings. The target sequence length in this task is
about 12.3 on average. Since this dataset consists of a parallel corpus of code snippets and docstrings, it is suitable for either the text-to-code retrieval task or the code summarization task.

To measure the prediction performance, we follow \cite{Alon2019} to use the BLEU score as the metric. For the baselines, we present results compared to 2-layer bidirectional LSTMs, Transformer, and Code2seq~\cite{alon2018code2seq}, a state-of-the-art model for code summarization task. We provide a fair comparison by splitting
tokens into subtokens and replacing UNK during inference. We also include numbers from the baselines used by~\citet{iyer2016summarizing}, such as MOSES~\cite{koehn2007moses} and an IR-based approach that use Levenshtein distance to retrieve the description.

\subsubsection{Results}
\vspace{-5pt}
Table~\ref{tab:code_summarization} shows the performance of Corder pretraining on the code summarization task. As seen, pre-training the model with the Corder pretext task outperform the other pre-training task, such as MNP or Masked LM in term of BLEU score with a significant margin for any of the encoder. 

\vspace{-5pt}
\subsection{Compared Against Supervised Methods}

An interesting question that one might ask is how Corder's performance in comparison with some supervised methods? The results for the tasks set out in the previous subsections may not be appropriate to answer this question because the three tasks are only used to measure how well the Corder Pretext task performs in a retrieval task (code-to-code retrieval) or in a fine-tuning process (text-to-code retrieval and code summarization), it is not clear if the Corder Pretext task can perform better than other supervised learning methods specifically trained for the tasks.
As such, a more appropriate {\it supervised} task is needed to answer this question.
A well-known evaluation protocol widely used to measure if the self-supervised learning methods can beat the supervised ones in natural language processing is to train the classifier (usually the Logistic Regression) on top of the embeddings provided by \textit{trained encoders on a supervised task} for classification tasks~\cite{kiros2015skip, logeswaran2018efficient}, and Accuracy is used as a proxy for representation quality. 
We adopt the \textit{code classification} task on the  OJ dataset~\cite{mou2016convolutional} and follow the similar evaluation protocol from \citet{logeswaran2018efficient} to evaluate if the embeddings provided by the self-supervised learning techniques are better than the supervised ones. The code classification task is to, given a piece of code, classify the functionality class it belongs to. The term \textit{code classification} is sometimes used interchangeably with the \textit{code summarization} task~\cite{sui2020flow2vec}, in which we want to automatically assign a label (or description) to a code snippet; it is just the way in which the label is generated can be different.

We produce the embeddings from all of the encoders for all of the training samples, then we train a classifier on top of these embeddings. We adapt a few strong state-of-the-art learning techniques for sentence representations in NLP to model the source code as a sequence of tokens, which are AdaSent~\cite{zhao2015self}, InferSent~\cite{conneau2017supervised}. The encoders from these two techniques are trained from supervised tasks.
We also use the encoder from the pretrained Code2vec on the method name prediction task as another baseline.
The embeddings produced by these techniques will also be used to train the multi-class classifier. We choose Logistic Regression as the classifier in this study. We use Accuracy as the metric to measure the performance of the code classification task. Table~\ref{tab:compared_supervised} shows the results of this analysis. The performance of the encoders trained from the Corder pretext performs significantly better than the other supervised learning methods.

\begin{table}[t]
	
	\centering
	\caption{Comparison against supervised representation learning methods on code classification task.}
	\label{tab:compared_supervised}
	\fontsize{8.4}{9.5}\selectfont 
	\begin{tabular}{|c|c|}
		\hline
		\textbf{Methods}           & \textbf{Accuracy}           \\ \hline
		\multicolumn{2}{|c|}{\textit{Pretrained from Supervised Methods}} \\ \hline
		AdaSent                    & 0.64                        \\ \hline
		InferSent                  & 0.56                        \\ \hline
		Code2vec                   & 0.75                        \\ \hline
		\multicolumn{2}{|c|}{\textit{Pretrained from Corder Pretext}}     \\ \hline
		Code2vec                   & 0.82                        \\ \hline
		TBCNN                      & 0.85                        \\ \hline
		Bi-LSTM                    & 0.76                        \\ \hline
		Transformer                & 0.80                        \\ \hline
	\end{tabular}
\end{table}

\subsection{Summary \& Threats to Validity}
\n{Added this summary section, need polishing}

Corder outperforms most of the baselines across three tasks: code-to-code retrieval, text-to-code retrieval, and code summarization. We also show that the embeddings produced by Corder perform better than the embeddings produced by the other supervised learning methods with code classification.

It should be noted that we try our best to use different pre-training tasks from other techniques to ensure comprehensive comparisons, but it is not easy to 
adapt all of the pretext tasks for all of the encoders, e.g., MNP in Code2vec/Code2seq, Mask LM in Transformer. It is because MNP is designed specifically for Code2vec/Code2seq. Adapting TBCNN/Bi-LSTM for MNP-pretraining depends on various factors and requires certain configurations such as choosing an AST parser, processing paths in ASTs, choosing the right parameters for the prediction layer, etc. Note that the choice of AST parsers alone could affect the performance of programming-language processing tasks significantly~\cite{bui2020treecaps}. There is no guarantee of the best settings for completely unbiased comparisons if we adapt. The same reasons are applicable to Masked-LM, which is designed specifically for the Transformer to process sequences. Thus, we only chose to adapt the pre-training task designed specifically for an encoder (MNP for Code2vec, Masked-LM for Transformer).

%% file: analysis.tex

We perform some analysis and ablation studies to measure how different design choices can affect the performance of Corder.

\begin{table}[t]
	\centering
	\caption{Results on Analysis on the Impact of Different Transformation Operators. TTC = Text-to-Code Retrieval (MRR as the metric), CS = Code Summarization (BLEU as the metric) }
	\label{tab:ablation_operator}
	\fontsize{7.7}{8.7}\selectfont 
	\begin{tabular}{|c|c|c|c|c|c|}
		\hline
		\multirow{2}{*}{\textbf{Models}} & \multirow{2}{*}{\textbf{Ops}} & \multicolumn{2}{c|}{\textbf{Original}} & \multicolumn{2}{c|}{\textbf{Downsampled }} \\ \cline{3-6} 
		&                               & \textbf{TTC}       & \textbf{CS}       & \textbf{TTC}            & \textbf{CS}           \\ \hline \hline
		\multirow{5}{*}{Code2vec}        & VR                            & 0.434              & 19.56             & 0.202                   & 16.78                 \\ \cline{2-6} 
		& US                            & 0.498              & 21.45             & 0345                    & 19.06                 \\ \cline{2-6} 
		& PS                            & 0.552              & 24.22             & 0.385                   & 20.01                 \\ \cline{2-6} 
		& SF                            & 0.401              & 19.11             & 0.401                   & 19.11                 \\ \cline{2-6}
		& LX                            & 0.423              & 21.43             & 0.320                   & 20.18                 \\ \cline{2-6}  
		& All                           & 0.592              & 26.56             & 0.419                   & 21.25                 \\ \hline
		\multirow{5}{*}{TBCNN}           & VR                            & 0.421              & 20.11             & 0.246                   & 17.89                 \\ \cline{2-6} 
		& US                            & 0.562              & 23.56             & 0.368                   & 18.24                 \\ \cline{2-6} 
		& PS                            & 0.603              & 22.98             & 0.320                   & 19.50                 \\ \cline{2-6} 
		& LX                            & 0.519              & 22.20             & 0.311                   & 19.82                 \\ \cline{2-6} 
		& SF                            & 0.354              & 18.75             & 0.461                   & 18.75                 \\ \cline{2-6} 
		& All                           & 0.702              & 25.39             & 0.398                   & 21.05                 \\ \hline
		\multirow{5}{*}{Bi-LSTM}         & VR                            & 0.423              & 21.53             & 0.302                   & 17.33                 \\ \cline{2-6} 
		& US                            & 0.601              & 22.42             & 0.401                   & 18.32                 \\ \cline{2-6} 
		& PS                            & 0.621              & 23.57             & 0.398                   & 19.56                 \\ \cline{2-6} 
		& LX                           & 0.529              & 20.89             & 0.328                   & 20.22                 \\ \cline{2-6} 
		& SF                            & 0.412              & 19.34             & 0.412                   & 19.34                 \\ \cline{2-6} 
		& All                           & 0.661              & 25.50             & 0.435                   & 20.84                 \\ \hline
		\multirow{5}{*}{Transformer}     & VR                            & 0.411              & 20.57             & 0.286                   & 17.82                 \\ \cline{2-6} 
		& US                            & 0.581              & 23.56             & 0.399                   & 19.45                 \\ \cline{2-6} 
		& PS                            & 0.639              & 24.12             & 0.410                   & 19.53                 \\ \cline{2-6} 
		& LX                            & 0.551              & 21.29             & 0.403                   & 19.77                 \\ \cline{2-6} 
		& SF                            & 0.403              & 18.10             & 0.418                   & 18.10                 \\ \cline{2-6} 
		& All                           & 0.728              & 26.41             & 0.440                   & 19.98                 \\ \hline
	\end{tabular}
\end{table}

\vspace{-5pt}
\subsection{Impact of Transformation Operators}
\label{impact}

We carry out an ablation study to evaluate how each transformation
operator affects the performance of particular code learning tasks. 
We perform separate training of our Corder pretext task using different transformation operators in our Corder training algorithm. In particular, when taking the transformation operators to transform the code snippet, the set of available operators $\mathcal{T}$ only contain one single operator. 

An issue with the comparison is that the number of code snippets per operator may vary, leading to unfair comparisons between the operators. This is because an operator is applied to modify the syntax and semantic of the code based on certain constraints, but such constraints may not apply all the time in the code. For example, the SF requires the snippet to contain at least one switch statement, but not all the snippets contain the switch statement. On the other hand, most of the snippets are applicable for VR because it is easy to change the variable names.
Concretely, for JavaMed, the number of snippets applicable for VR, PS, US, LX, SF, are 919823, 64953, 352443, 26042, 3213, respectively. To see the impact of the operators, we perform this analysis under two settings: (1) we perform the transformation for each of the operators on the original number of snippets; and (2) we only select 3213 snippets for VR, PS, and US, in which the snippets must be the same as the snippets applicable for SF.
We train Corder on the encoders with similar settings in the Evaluation
Section, but we only use one operator at a time. 
Then we perform the fine-tuning process on the text-to-code retrieval and code summarization task, also similar to the Evaluation Section. The results of this analysis can be seen in Table~\ref{tab:ablation_operator}. The column "Original" means the analysis is performed when using all of the snippets. The columns "Downsampled" means the analysis is performed when downsampling the number of snippets for VR, US, PS into the same number of snippets of SF.
There are a few observations:
\begin{itemize}[leftmargin=*]
	\item In the "Original" setting, although VR has the most number of applicable snippets, its performance is among the worst, for either Text-to-Code Retrieval or Code Summarization. PS and US are two operators that perform the best (PS is slightly better most of the time). The performance of SF is comparable to VR but the number of snippets for SF is much fewer.  
	\item In the "Downsampled" setting, SF becomes as comparable to PS and US, and these 3 operators perform much better than the VR, either in Text-to-Code Retrieval or Code Summarization. SF also performs the best in Text-to-Code Retrieval.
\end{itemize}
With the observations, we can conclude that changing the code structures is crucial to learn a decent model of source code. VR only changes the text information, while PS, US, and SF modify the code structure extensively. 

\subsection{Visualization for Code Embeddings}

\begin{figure}[t]
	\centering
	\begin{tabular}{@{}cccc@{}}		
		\includegraphics[width=0.7\columnwidth]{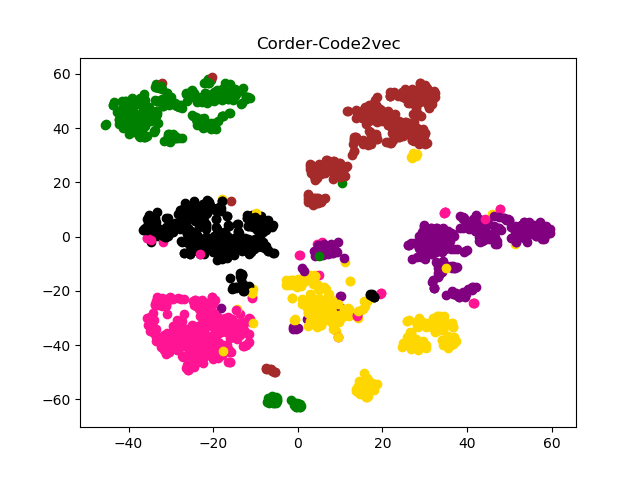} \\
		\includegraphics[width=0.7\columnwidth]{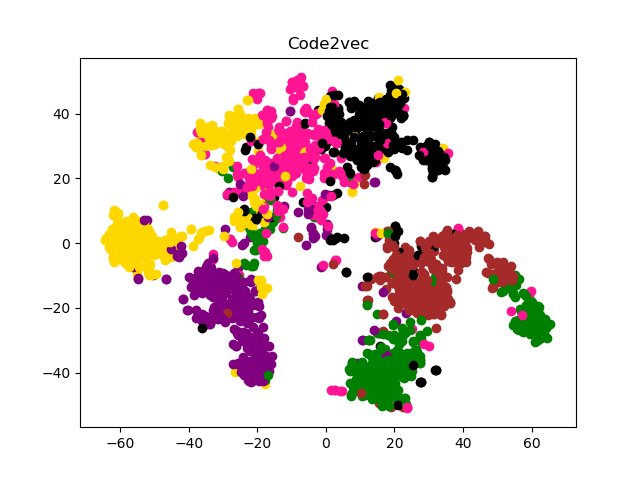}
	\end{tabular}
	\vspace*{-10pt}
	\caption{Visualization of the vector representations of the code snippets from 6 classes in the OJ Dataset produced by Corder-Code2vec and MNP-Code2vec}
	\label{fig:visualization}
\end{figure}

We visualize the code vectors to help understand and explain why the vectors produced by Corder pre-training are better than the vectors produced by other
We choose Code2vec as the encoder for this analysis since Code2vec has been adapted in two different pretext tasks: (1) Corder pretext task (Corder-Code2vec); and (2) method name prediction task~\cite{Alon2019} (Code2vec).
The goal is to show that our Corder pretext task performs better than the method name prediction as a pretext task to train the source code model. We use the code snippets in OJ dataset~\cite{mou2016convolutional} that has been used for the code-to-code retrieval task. We randomly choose the embeddings of the first 6 classes of the OJ dataset then we use T-SNE~\cite{maaten2008visualizing} to reduce the dimensionality of the vectors into two-dimensional space and visualize. As shown in Figure~\ref{fig:visualization}, the vectors produced by Corder-Code2vec group similar code snippets into the same cluster with much clearer boundaries. This means that our instance discrimination task is a better pretext task than the method name prediction task in ~\citet{Alon2019} for the same Code2vec encoder.

%% file: conclusion.tex
We have proposed Corder, a self-supervised learning approach that can leverage large-scale unlabeled data of source code. Corder works by training the neural network over a contrastive learning objective to compare similar and dissimilar code snippets that are generated from a set of program transformation operators. The snippets produced by such operators are syntactically diverse but semantically equivalent.
The goal of the contrastive learning method is to minimize the distance between the representations of similar snippets (positives) and maximize the distance between dissimilar snippets (negatives). We have adapted Corder for three tasks: code-to-code retrieval, fine-tuning for text-to-code retrieval, fine-tuning for code summarization, and have found that Corder pre-training significantly outperforms other models not using contrastive learning
on these three tasks.

%% file: ack.tex
This research is supported by the Singapore Ministry of Education (MOE) Academic Research Fund (AcRF) Tier 2 Award No.~MOE2019-T2-1-193 and RISE Lab Operational Fund from SCIS at SMU, Royal Society projects (IES/R1/191138, IES/R3/193175), EPSRC STRIDE project (EP/T017465/1), and Huawei Trustworthy Software Engineering and Open Source Lab. We also thank the anonymous reviewers for their insightful comments and suggestions, and thank the authors of related work for sharing data.